\documentclass{JINST}
\usepackage{graphicx}

\title{From a single encapsulated detector to the spectrometer for INTEGRAL satellite: predicting the peak-to-total ratio at high $\gamma$-energies}

\author{Ritesh Kshetri\thanks{Present address: INFN Laboratori Nazionali del Sud, via S.Sofia 62, 95125 Catania, Italy.}\\
  Nuclear Physics Division, Saha Institute of Nuclear Physics,\\
  1/AF, Bidhannagar, Kolkata - 700064, India.\\

  E-mail: \email{ritesh.khetri@saha.ac.in}}

\abstract{In two recent papers (R. Kshetri, JINST 2012 7 P04008; ibid., P07006), a probabilistic formalism was introduced to predict the response of encapsulated type composite germanium detectors like the SPI (spectrometer for INTEGRAL satellite). Predictions for the peak-to-total and peak-to-background ratios are given at 1.3 MeV for the addback mode of operation. The application of the formalism to clover germanium detector is discussed in two separate papers (R. Kshetri, JINST 2012 7 P07008; ibid., P08015). Using the basic approach developed in those papers, for the first time we present a procedure for calculating the peak-to-total ratio of the cluster detector for $\gamma$-energies up to 8 MeV. Results are shown for both bare and suppressed detectors as well as for the single crystal and addback modes of operation. We have considered the experimental data of (i) peak-to-total ratio at 1.3 MeV, and (ii) single detector efficiency and addback factor for other energies up to 8 MeV. Using this data, an approximate method of calculating the peak-to-total ratio of other composite detectors, is shown. Experimental validation of our approach (for energies up to 8 MeV) has been confirmed considering the data of the SPI spectrometer. We have discussed about comparisons between various modes of operation and suppression cases. The present paper is the fifth in the series of papers on composite germanium detectors and for the first time discusses about the change in fold distribution and peak-to-total ratio for sophisticated detectors consisting of several modules of miniball, cluster and SPI detectors. Our work could provide a guidance in designing new composite detectors and in performing experimental studies with the existing detectors for high energy gamma-rays.}


\dedicated{Dedicated to my late wife Manjari}

\begin{document}

\section{Introduction}

High-purity germanium (HPGe) detectors have excellent energy resolution and reasonable full energy peak (FEP) efficiency for high resolution gamma-ray spectroscopy \cite{ge}. Higher FEP efficiency requires large sized HPGe detectors having large active volume. However, larger crystals are difficult to grow and have a larger Doppler broadening for gamma-rays emitted from nuclei in motion, which deteriorates energy resolution. A way of obtaining high detection efficiency without compromising energy resolution is the use of encapsulated type composite detectors like the miniball detector array and the cluster detector. The former comprises of three and four detector modules packed inside two types of cryostats \cite{ge,ebe}, while the latter consist of seven hexagonal encapsulated HPGe detector modules (kept inside the same cryostat) \cite{ge,ebe,ebe2}. There are also composite detectors without encapsulated modules like the clover germanium detector \cite{clov} where four HPGe crystals are kept inside the same cryostat. For more than a decade, these composite detectors have played a pivotal role in in-beam and decay spectroscopy experiments \cite{ge}.

A similar detector is the SPI spectrometer (spectrometer for INTEGRAL satellite) \cite{spi2}, which addresses the gamma spectroscopy of celestial $\gamma$-ray sources. It consists of an array of nineteen closely packed encapsulated HPGe detectors surrounded by an active anticoincidence shield of bismuth germanate \cite{spi2}. Figure 1(Ai) shows the three detector module cryostat of the miniball array while figures 1(Bi) and 1(Ci) show the cluster detector and the SPI spectrometer, respectively. Here, {\it K} is the number of constituent modules of a composite detector. The schematic diagrams are shown beside the respective figures. Comparing figures 1(Bi) and 1(Ci), we observe that the SPI spectrometer comprises of twelve detector modules surrounding the cluster detector. It would be interesting to compare the fold distribution and peak-to-total ratio of these three detectors with composite detectors comprising of arrays of these detectors. Figures 1(Aii), 1(Bii) and 1(Cii) comprise of three modules of miniball, cluster and SPI detectors. Similarly, figures 1(Aiii), 1(Biii) and 1(Ciii) comprise of seven modules of miniball, cluster and SPI detectors. 
\newline

\begin{figure}[htp]
\centering
\includegraphics[totalheight=0.495\textheight,viewport=75 260 745 795,clip]{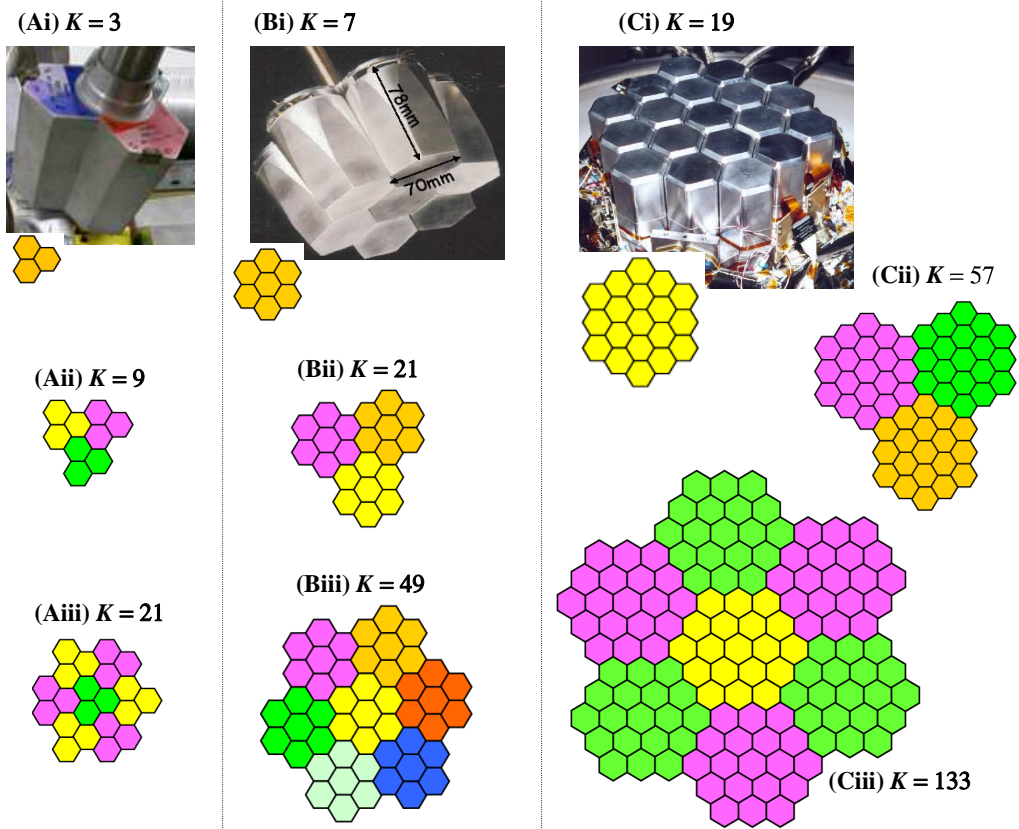}
\caption{Composite detectors including the three detector module cryostat of the miniball array, the cluster detector, the SPI spectrometer are shown along with more sophisticated detectors comprising of arrays of these three detectors.}
\end{figure}

For $\gamma$-ray spectrometers like the composite detectors, knowledge about the FEP efficiency and the total efficiency are important measurable quantities. In case of the total efficiency, all interactions of incident $\gamma$-ray with detector, are assumed to be counted, while for peak efficiency, only those interactions that deposit the full energy of the $\gamma$-ray, are counted. Both of these efficiencies contain the dependence on the source-to-detector distance. So, it is more convenient to use the peak-to-total ratio (the ratio between the peak and total efficiencies) than the total efficiency itself, since the spatial dependence is smoother for this ratio as it is expected to cancel out. As a result, the influence of the intrinsic properties of the detector on the total-to-peak ratio is more pronounced than on the total efficiency or peak efficiency. In the literature, there has been numerous studies regarding the peak efficiency, total efficiency and peak-to-total ratio for single HPGe detectors. Some of the works could be found in ref. \cite{pt}. For composite detectors, there have been several simulation studies of the peak-to-total ratio \cite{ebe2,clov,pt2,tig1}.

In two recent papers \cite{jinst,jinst2}, we have developed a phenomenological approach of understanding the operation of various composite detectors in addback mode. From the basic considerations of absorption and scattering of gamma-rays, a probabilistic formalism has been presented such that a general composite detector could be described in terms of four quantities only. Calculations of the peak-to-total ratio have been performed for several detectors including the miniball, cluster and SPI detectors. One can question how the detailed physics of the various interaction processes are incorporated in our formalism. It is noteworthy that for gamma-rays with energies from 10 keV to 10 MeV, three types of interactions are important for gamma-ray detection: the photoelectric absorption, the Compton scattering and the pair production (possible when gamma-ray energy $\ge$ 1.022 MeV). Our formalism includes the detailed physics of these interaction processes through the experimental inputs of (i) peak-to-total ratio at 1.3 MeV, and (ii) single detector efficiency and addback factor for the gamma energy of interest. This fact is supported by the agreement between the theoretical results and the experimental data \cite{jinst,jinst2}.

Instead of using an empirical method or simulation, our works \cite{jinst,jinst2} present the first unified approach to calculate the peak-to-total ratio using experimental data as input and presents another way of understanding the operation of composite detectors. In those works \cite{jinst,jinst2}, we showed that using the experimental data of cluster detector at 1.3 MeV, the formalism could be used to predict the peak-to-total ratio as a function of number of detector modules.  Similar approaches for modeling the clover detector are presented in two recent papers \cite{jinst3,jinst4}. Remarkable agreement between experimental data and analysis results has been observed for composite detectors like TIGRESS clover detector and SPI spectrometer at 1.3 MeV \cite{jinst2,jinst3}.

The present paper is the fifth in the series of papers on composite germanium detectors where the formalism has been further developed starting with the generalized approach discussed in section 2 of ref. \cite{jinst2}. The investigations of this paper that differentiates this manuscript from its predecessors (see \cite{jinst,jinst2,jinst3,jinst4}) are pointed below:

\begin{itemize} 

\item We have performed the calculations of the peak-to-total ratio of the cluster detector for higher energies up to 8 MeV using the experimental data. We have also shown an approximate method of calculating the addback factor and the peak-to-total ratio of the composite detectors. Using the present approach, for the first time all single HPGe and composite detectors could be described in a single framework for various energies. 
\footnote{It is noteworthy that significant work has been done in the past two decades on more sophisticated detectors including highly segmented, position-sensitive and tracking HPGe detectors for a variety of applications \cite{ge}. In this context, a phenomenological approach for the composite detectors could seem to be outdated. However, even if these detectors were developed more than a decade ago, still it does not decrease the significance of this work which attempts to undertake from first principles the analysis of a complex detection problem.}

\item Both bare and suppressed detectors are considered for the composite detector working in single detector and addback modes of operation. Based on background counts, we have discussed about the comparisons between the various modes of operation and suppression cases.

\item Regarding the composite detectors, one could ask what would be the increase in performance if we compare the SPI spectrometer with an array of several SPI spectrometers. In order to answer this question, for the first time we have discussed about the change in these performance figures for sophisticated detectors consisting of several modules of miniball, cluster and SPI detectors. Results are shown for higher energies upto 8 MeV.

\end{itemize}

\section{Understanding the operation of gamma detectors}

\subsection{A probabilistic approach for single detector}

We will first consider a single HPGe detector having identical shape, size and volume as one of the detector modules of the cluster detector.
\footnote{A comparison of relative efficiency for addback mode of cluster detector with a 30$\%$ HPGe detector for gamma energies up to 10 MeV could be found in \cite{wil}.}
Let $N_T$ be the total flux of monoenergetic $\gamma$-rays (of energy $E_{\gamma}$) incident on the detector such that at a time a single $\gamma$-ray could interact. Let $N$ be the portion of the total flux that interact with the detector. If the probability of FEP absorption be $A$, then $NA$ $\gamma$-rays are fully absorbed and $N(1 - A)$ $\gamma$-rays escape from the detector after partial energy deposition, thereby contributing to background. Let the full energy peak, background and total counts be denoted by $p$, $b$ and $t$ respectively. Also, there could be background counts due to 

\begin{itemize}

\item $\gamma$-rays which interact with nearby passive material including suppression shield and afterwards interact with the detector, and 

\item $\gamma$-rays which escape the detector after partial energy deposition. Some of them undergo scattering from surrounding materials and afterwards interact again with the detector. 

\end{itemize}
Due to these cases, let the number of background counts be $N\eta$, where $\eta$ is a parameter. So, total background counts are $N(1 - A + \eta)$. The peak-to-total ratio of this bare detector is given by
\begin{equation}
p/t = \frac{A}{1 + \eta}
\end{equation}
For suppressed detector (with active suppression), if events corresponding to \{$N(1 - A)$\} and $N\eta$ get subtracted by a fraction $\kappa$ (namely suppression factor), then there is a reduction in background counts, given by $[b - \kappa \{N(1 - A) + N\eta\}]$. Total counts are given by $t^S = N(1 + \eta) - \kappa N\{(1 - A) + \eta\}$. The peak-to-total ratio is given by
\begin{equation}
p/t^S = \frac{A}{(1 + \eta) - \kappa\{(1 - A) + \eta\}}
\end{equation}
Comparing equations 2.1 and 2.2, we have $p/t^S > p/t$.

\subsection{Basic principle of operation of a composite detector}

If a $\gamma$-ray is incident on a composite detector (consisting of {\it K} hexagonal encapsulated HPGe detector modules), then it can interact with one of the detector modules after which the $\gamma$-ray can be absorbed or scattered. For the latter case, there is partial absorption of energy in the crystal of interaction (or modules of interaction). The scattered $\gamma$-ray from one crystal can enter adjacent crystal (without undergoing energy loss from the material between the crystals) where it can again interact (be absorbed / undergo scattering). There is also the possibility of the scattered $\gamma$-ray escaping the detector system. The interaction of a $\gamma$-ray with several crystals can cause time correlated events and if we add up these time correlated events, then information from the scattered $\gamma$-rays which do not escape from the detector system, will be added back to the full energy peak (FEP), i.e, we can get back the full energy of incident $\gamma$-ray that undergoes scattering(s) after partial energy absorption. Thus, the events involving scattered $\gamma$-ray(s) which do not escape detection are reconstructed. The composite detector could be operated in two modes \cite{clov}:

\begin{itemize}

\item single detector mode which is the time uncorrelated sum of up to $K$ crystal data corresponding to events where the full $\gamma$-ray energy is  deposited in any one of the individual modules, 

\item addback mode which is the time correlated sum of up to $K$ crystal data. The later mode corresponds to events where the full $\gamma$-ray energy is deposited by single and multiple hits. 

\end{itemize}

\noindent Due to the multiple hit events, there are more peak counts and less background (including escape peaks) counts in addback mode. As a result, the efficiency and peak-to-total ratio are higher in addback mode compared to single crystal mode. An illustration of operation of cluster detector is shown in figure 2, which shows five ways in which a 1 MeV gamma-ray could deposit its energy in a cluster detector. This figure is used to illustrate the difference between the single detector and addback modes of operation. The various cases are discussed below:
\newline

\begin{figure}[htp]
\centering
\includegraphics[totalheight=0.5\textheight,viewport=75 260 750 795,clip]{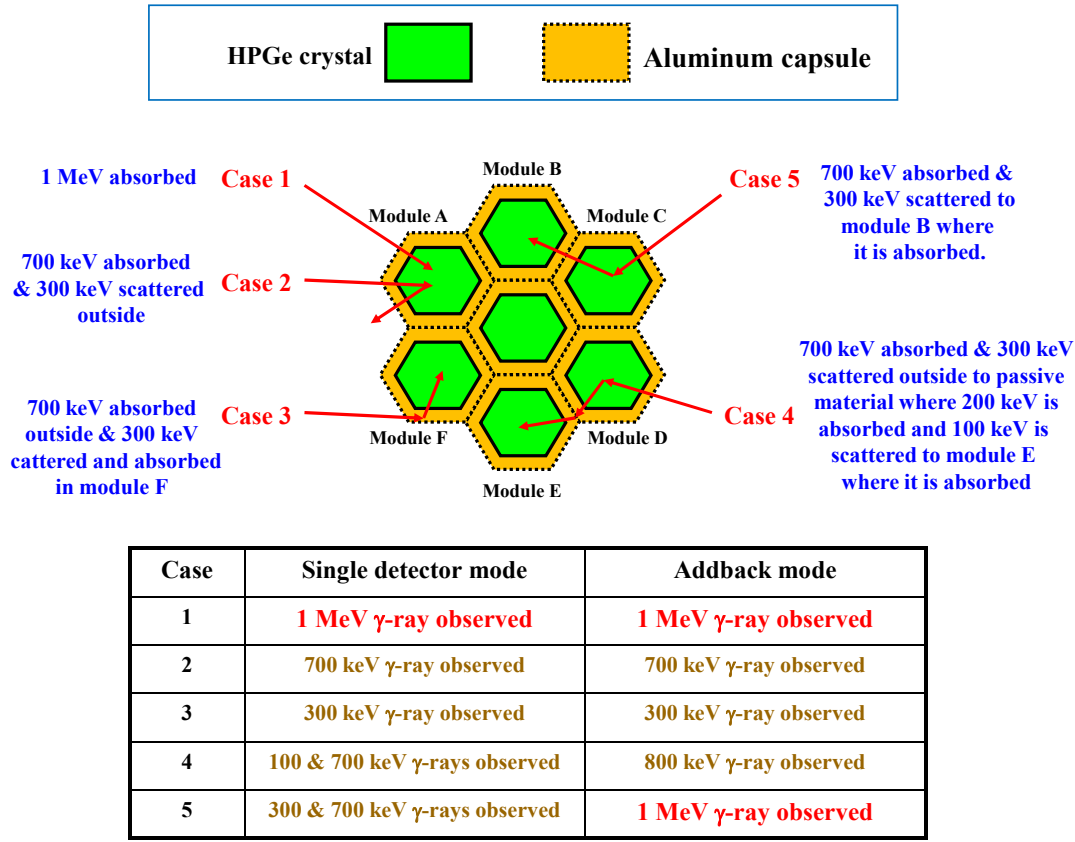}
\caption{Comparison of the two modes of operation of a bare cluster detector (without escape suppression shield). The schematic diagram shows the interaction of a gamma-ray of energy 1 MeV with the detector for five cases (see text for details).}
\end{figure}

\begin{enumerate}

\item The first case corresponds to complete absorption of gamma-ray after interaction with a single detector module and so this event is detected in both modes. 

\item Here, the gamma-ray undergoes Compton scattering where 700 keV energy is absorbed in HPGe crystal of module A. The Compton scattered gamma-ray of energy 300 keV escapes from the module. For both modes, a 700 keV $\gamma$-ray (contributing to background) is detected. 

\item The gamma-ray undergoes Compton scattering where 700 keV energy is absorbed in passive material surrounding the modules. Compton scattered gamma-ray of energy 300 keV enters module B where it is absorbed in the HPGe crystal. For both modes, a 300 keV $\gamma$-ray (contributing to background) is detected.

\item In the fourth case, the gamma-ray undergoes Compton scattering where only 700 keV energy is absorbed in HPGe crystal of module D and 300 keV is scattered outside. Due to interaction with the passive material, 200 keV is absorbed and Compton scattered gamma-ray of energy 100 keV enters the module C where it is absorbed. This event, in the single crystal mode, will be registered as two events of energies 100 keV and 700 keV. While in addback mode, these two time correlated events are added back and a single event of energy 800 keV is observed. However, for both cases, the event contributes to background.

\item The Compton scattered gamma-ray of energy 300 keV escapes from the module E and is detected in module F where it is completely absorbed. This event, in the single crystal mode, will be registered as two events of energies 300 keV and 700 keV. While in addback mode, these two time correlated events are added back and a single event of energy 1 MeV will be observed. 

\end{enumerate}

\noindent In the first and fifth cases, the $\gamma$-ray deposits its energy completely in the detector. For the rest, there is incomplete energy deposition. In the first case, we observe a single count in single crystal spectrum (this FEP event is called single fold event). The fifth case is an example of multiple fold event where $m$ counts ($2 \le m \le 7$) are observed in single crystal spectrum (this FEP event is called multiple fold event). Thus, we observe that a FEP event can generate $m$ background counts in spectrum for single crystal mode of operation. However, in the spectrum corresponding to addback mode, the $m$ background counts are reconstructed to a single FEP count.

\subsection{A probabilistic approach for composite detector}

Let $N_T$ be the total flux of monoenergetic $\gamma$-rays (of energy $E_{\gamma}$) incident on a composite detector (with $K$ modules; $K$ = 7 for cluster and 19 for SPI) and $N$ be the portion of the total flux that interact with a detector module of the cluster detector, such that $NK$ $\gamma$-rays interact with the composite detector. We assume that at a time a single $\gamma$-ray interacts with one of the seven detector modules. After interaction, we have the following possibilities:

\begin{itemize}

\item Some of the gamma-rays are fully absorbed in a single detector module. Let the probability of FEP absorption after single detector interaction be $A$. 

\item Some gamma-rays are fully absorbed after interaction with two or more detector modules. Let the probability of FEP absorption after multiple detector interaction be $\alpha$. 

\item Some gamma-rays escape the composite detector leading to partial energy absorption. The gamma-rays could escape after interaction with one or more detector module(s). The events corresponding to these gamma-rays will contribute to background. Let the probability of scattering away from the detector be $\beta$.

\end{itemize}

\noindent So, we have, $A + \alpha + \beta =$ 1. $A, \alpha$ and $\beta$ are probabilities integrated over energies and angles of scattered gamma-rays. Apart from the above mentioned cases, there could be two additional sources of background counts ($N\eta$) discussed in previous section 2.1. So, total background counts are $N(\beta + \eta)$. Let the full energy peak, background and total counts in addback mode be denoted by $P$, $B$ and $T$ respectively. The total counts are given by
\begin{equation}
T = NK + NK\eta 
\end{equation}
\begin{equation}
 = NKA + NK\alpha + NK\beta + NK\eta
\end{equation}

\subsubsection{Single detector mode of operation}

In single detector mode, the FEP events are $NKA$ and the events contributing to background are 
\begin{enumerate}

\item the FEP events where the gamma energy is deposited in several crystals,
\footnote{As mentioned before, a FEP event where the gamma energy is deposited in $m$ crystals, could generate $m$ background counts in spectrum for single crystal mode of operation.}
and 

\item the events where $\gamma$-rays escape the HPGe volume after partial energy deposition in a single or several crystals. 

\end{enumerate}
The multiple scattering events corresponding to case 1 above, where the partial energies released are not summed, account for ($N\alpha + N\alpha_1$) counts which contribute to the background. Regarding case 2, the number of events where $\gamma$-rays escape from the cluster detector in addback mode is $N\beta$. Let us consider an example where a $\gamma$-ray escapes the detector after partial energy deposition in two crystals. In spectrum for addback mode, a single count is observed. However, for single crystal mode, two counts are observed. Thus, counts corresponding to escaping events are greater than $N\beta$ in single crystal mode. Let the number of such counts be given by ($N\beta + N\beta_1$). Thus, the peak-to-total ratio in single crystal mode for a bare composite detector is given by $\frac{A}{1 + \alpha_1 + \beta_1 + \eta}$.

\subsubsection{Addback mode of operation}

For the addback mode, we have
\begin{equation}
P = NKA + NK\alpha  ,   
\end{equation}    
\begin{equation}
B = NK(\beta + \eta)
\end{equation}
Using $A + \alpha + \beta =$ 1, the peak-to-total ratio is given by
\begin{equation}
P/T = \frac{A + \alpha}{1 + \eta}   
\end{equation}
Comparing with equation 2.1 and peak-to-total ratio for single detector mode (section 2.3.1), we observe that the peak-to-total ratio in single detector mode is lower than that of a single HPGe detector (having identical shape, size and volume as one of the detector modules of the composite detector). Note that the peak-to-total ratio in addback mode has the highest value. Thus, we have $(P/T)_{sd} < (P/T)_{single~HPGe} < (P/T)_{addback}$.

The comparison of detector performance in addback and single detector modes is given by the addback factor ($F$), which is ratio of the relative efficiency in addback mode ($\epsilon_{adbk}$) to that in single detector mode ($\epsilon_{sd}$)
\footnote{Note that $\epsilon_{sd} \propto A$ and $\epsilon_{adbk} \propto (A + \alpha)$}
, given by
\begin{eqnarray}
F &=& \frac{\epsilon_{adbk}}{\epsilon_{sd}} \nonumber \\
  &=& \frac{A + \alpha}{A} = 1 + \frac{\alpha}{A}  
\end{eqnarray} 
It is observed that the knowledge of $A$ and $\alpha$ results in determination of peak-to-total ratio and addback factor for various values of $\eta$. If the number of single and multiple detector events contributing to FEP be $NKh_s$ and $NKh_m$, then
\begin{equation}
h_s = \frac{A}{A + \alpha}  , 
\end{equation} 
\begin{equation}
h_m = \frac{\alpha}{A + \alpha}   
\end{equation} 
The variation of $h_s$ and $h_m$ as a function of gamma energy gives us information about the fold distribution of a composite detector. Note that $h_s = \frac{1}{F}$.
\newline

\noindent {\bf A suppressed detector in addback mode:}
If events escaping the detector ($N \beta$) and events corresponding to $N\eta$, get subtracted by a fraction $\kappa$ (namely suppression factor), then there is a reduction in background counts. If the background and total counts be denoted by $B^S$ and $T^S$, respectively. Then, we have 
\begin{equation}
B^S = B - NK\kappa(\beta + \eta)
\end{equation}
So, total counts are given by   
\[ T^S = (NK + NK\eta) - NK\kappa(\beta + \eta) \]
\begin{equation}
       = NK(1 + \eta) - NK\kappa(\beta + \eta)
\end{equation}
With increasing value of $\kappa$ $(0 \le \kappa \le 1)$, we have better active suppression. Passive suppression corresponds to $\kappa = 0$. Thus, the peak-to-total ratio (in addback mode) for a suppressed detector is given by
\begin{equation}
P/T^S = \frac{A + \alpha}{1 + \eta - \kappa(\beta + \eta)} .
\end{equation}       
The expressions for addback factor and peak-to-total ratio (bare and suppressed cases) for addback mode depend on $A'$, $\alpha$ and are independent of number of detector modules. These expressions are valid for any composite detector. However, the values will differ depending on values of $A, \alpha, \kappa$ and $\eta$. It is noteworthy that the peak-to-total ratio (equation 2.13) for $\eta = \kappa = 0$ and the relative FEP efficiency ($\propto (A + \alpha)$) are identical. Increasing the value of $\eta$ increases the background, thereby decreasing the value of peak-to-total ratio. Similarly, if $\kappa$ increases, background decreases and thus the  peak-to-total ratio increases.

\subsection{Observations from previous works}

In our recent work \cite{jinst2}, we showed that for a general composite detector with {\it K} hexagonal detector modules, the probability amplitude $\alpha$ for a monoenergetic $\gamma$-ray, could be expressed (see equation 3.2 of ref. \cite{jinst2}) as
\begin{equation}
\alpha = \rho (S'_1A) + \rho^2 S_1(S'_1A)
\end{equation} 
where the coefficient $\rho$ is the ratio of covered lateral surface area to total lateral surface area of the composite detector, and $A, S_1, S'_1$ are probabilities describing the interaction of a monoenergetic $\gamma$-ray with a composite detector. Here, the first and second terms correspond to double and triple hit events which contribute to FEP in addback mode. If we denote $S'_1A$ by $x_1$ and $S_1S'_1A$ by $x_2$, then we have
\begin{equation}
\alpha = \rho x_1 + \rho^2 x_2    
\end{equation}  
Note that the knowledge of $A$ and $\alpha$ are sufficient to know $\beta$ since $A + \alpha + \beta = 1$. In order to find expressions for $\alpha_1$ and $\beta_1$, we can consider the detailed calculations for the clover detector as shown in section 2.3 of ref. \cite{jinst4}. Similarly, for the present case, it could be shown that
\begin{equation}
\alpha_1 = \rho x_1 + 2\rho^2 x_2    
\end{equation}  
For simplicity, we make two approximations: $\alpha_1 \approx \alpha$ and $\beta_1 = 0$. Thus, from the present formalism, it is possible get an estimate of the peak-to-total ratio in single detector mode given by
\begin{equation}
(P/T)_{sd} = \frac{A}{1 + \alpha + \eta} 
\end{equation}  
\newline

\noindent{\bf An interesting application:} From the above formulation, it is possible to know the value of $\alpha$ of a composite detector if we know the same for a different composite detector at a given energy. Let us consider the cluster detector and another composite detector, such that 
\begin{equation}
\alpha_{cluster} = \rho x_1 + \rho^2 x_2    
\end{equation} 
and
\begin{equation}
\alpha_{det} = \rho_0 x_1 + \rho_0^2 x_2    .
\end{equation} 
If we consider the quantity
\begin{equation}
\alpha'_{det} = \frac{\rho_0}{\rho} \alpha_{cluster} = \rho_0 x_1 + \rho_0 \rho x_2    ,
\end{equation} 
it is observed that $\alpha'_{det} \approx \alpha_{det}$. Thus, using $\alpha'_{det}$, we can find the peak-to-total ratio of a general composite detector, if we know the value of one of them (cluster detector in the present case) at an energy.

\subsection{Comparison of modes of operation and suppression}

\begin{itemize}

\item Peak counts increase and background counts decrease in addback spectrum compared to single detector spectrum. However, in case of suppression, only background counts decrease compared to that of a bare detector. In order to find which operation causes better performance (with more decrease in background counts), let us define a ratio ($R_{1}$) of decrease in background counts due to addback to the decrease in counts due to suppression, given by
\begin{eqnarray}
R_{1} &=& \frac{B_{sd} - B_{adbk}}{B_{sd} - B_{sd}^S} \nonumber \\
      &=& \frac{NK(\alpha + \beta + \alpha_1 + \beta_1 + \eta) - NK(\beta + \eta)}{NK(\alpha + \beta + \alpha_1 + \beta_1 + \eta) - NK[(\alpha + \beta + \alpha_1 + \beta_1 + \eta) - \kappa(\beta + \eta)]} \nonumber \\
      &=& \frac{2\alpha}{\kappa(\beta + \eta)} 
\end{eqnarray} 
Here we have considered the two approximations: $\alpha_1 \approx \alpha$ and $\beta_1 = 0$ in the last expression.

\item For comparing the performance of a suppressed detector with a bare one, let us consider the ratio ($R_{2}$) of background counts for suppressed case to that for bare case, given by
\begin{equation}
R_{2} = \frac{1}{2}[\frac{B_{sd}^S}{B_{sd}} + \frac{B_{adbk}^S}{B_{adbk}}] = 1 - \kappa \delta  . 
\end{equation} 
where $\delta = \frac{1}{2}[1 + \frac{\beta + \eta}{2\alpha + \beta + \eta}]$, considering $\alpha_1 \approx \alpha$ and $\beta_1 = 0$. Note that here we have considered both modes of operation of the cluster detector. Lower value of $R_2$ means better performance in active suppression. In case of ideal suppression ($\kappa = 1$), $R_{1}$ and $R_{2}$ attain respective minimum values of $(\frac{2\alpha}{\beta + \eta})$ and ($1 - \delta$). 

\end{itemize}

\section{Predictions for single HPGe and cluster detectors}

The theoretical formalism developed in the last section could be realized in practice and it is possible to predict the peak-to-total ratio for an energy (say $E_0$) where experimental data about this ratio is absent. For a composite detector, if we take as input for our formalism, the following known experimental data:
\begin{itemize}
\item single crystal FEP efficiency, addback factor at energy $E_0$, and
\item peak-to-total ratio, single crystal FEP efficiency, addback factor at a lower energy (say $E_1$),
\end{itemize}
then, it is possible to predict the peak-to-total ratio at $E_0$. Along with the extrapolation from known data at lower energies and the Monte Carlo simulations (e.g. GEANT) that are validated at lower energies, our formalism presents a way of knowing the peak-to-total ratio for energy regions with no information about them. The methodology will be illustrated in this section.

We will consider the experimental data for addback factor ($F$) and relative efficiency in single detector mode ($\epsilon_{adbk}$) of a cluster detector \cite{wil} for four $\gamma$-ray energies (E$_{\gamma} \approx$ 1.3, 3.8, 5.4 and 8.0 MeV). Approximate values of $F$ and $\epsilon_{adbk}$ (taken from ref. \cite{wil}) are shown in table 1.  From relative efficiency in addback mode and addback factor, the value of relative efficiency in single detector mode ($\epsilon_{sd}$) has been extracted (note that $F = \frac{\epsilon_{adbk}}{\epsilon_{sd}}$). From equations 2.7 and 2.8, we get
\begin{equation}
A = (1 + \eta)\frac{(P/T)}{F}
\end{equation} 
Let us assume $\eta$ = 0.1. The experimental value of peak-to-total ratio in addback mode at 1.3 MeV is found to be $\approx$ 0.39 (see table 1 of ref. \cite{ebe}). For 1.3 MeV $\gamma$-ray, substituting the values of addback factor (= 1.5, from table 1) and peak-to-total ratio (= 0.39), equation 3.1 gives $A = 0.29$. Now, $\frac{\epsilon_{rel}^{sc}}{A}$ = $\frac{0.33}{0.29} \approx 1.14$. So, for $E_{\gamma}$ = 3.8 MeV, if we divide $A$ (from table 1) by 1.14, we get the respective value of $A$ ( = 0.14). Using this, similar to previous case, the value of peak-to-total ratio is found to be $\frac{1.82\times0.14}{1 + 0.1} =$ 0.23. This procedure could be repeated for other energies. The peak-to-total ratio has been calculated as a function of $\gamma$-ray energy for an HPGe detector (having identical shape, size and volume as one of the detector modules of the cluster detector) and the two modes of operation of the cluster detector. For calculating peak-to-total in single detector mode, we have used equation 2.17. The results are shown in figure 3 for both bare and suppressed cases using $\eta = 0.1$. Similar to FEP efficiency, a decreasing trend with higher $\gamma$-energy is observed. The trend is similar for different suppression cases. Improvement due to active suppression is observed. Compared to that of single HPGe detector, the peak-to-total ratio of cluster is higher by almost 1.5 times for both bare and suppressed cases. The peak-to-total ratio for bare and suppressed cluster (with $\kappa$ = 0.5) detector at 8 MeV are 8 $\%$ and 14 $\%$, respectively.

\begin{table}
\begin{center}
\setlength{\tabcolsep}{0.03in} 
{\caption{\label{tab:table1} Approximate values of the relative efficiency in addback mode ($\epsilon_{adbk}$) and the addback factor ($F$) for four energies of a cluster detector \cite{wil}.}
\vspace{1.5ex}
\begin{tabular}{|c|p{1.2cm}|p{1.2cm}|p{1.2cm}|p{1.2cm}|} \hline
E$_\gamma$ (MeV) & 1.3 & 3.8 & 5.4 & 8.0 \\
\hline
$\epsilon_{adbk}$ & 0.5 & 0.3 & 0.2 & 0.1 \\
$F$ & 1.5 & 1.82 & 2.03 & 2.45 \\
$\epsilon_{sd}$ & 0.33 & 0.16 & 0.10 & 0.04 \\
\hline
\end{tabular}}
\normalsize
\end{center}
\end{table}

\begin{figure}[htp]
\centering
\includegraphics[totalheight=0.5\textheight,viewport=40 270 780 785,clip]{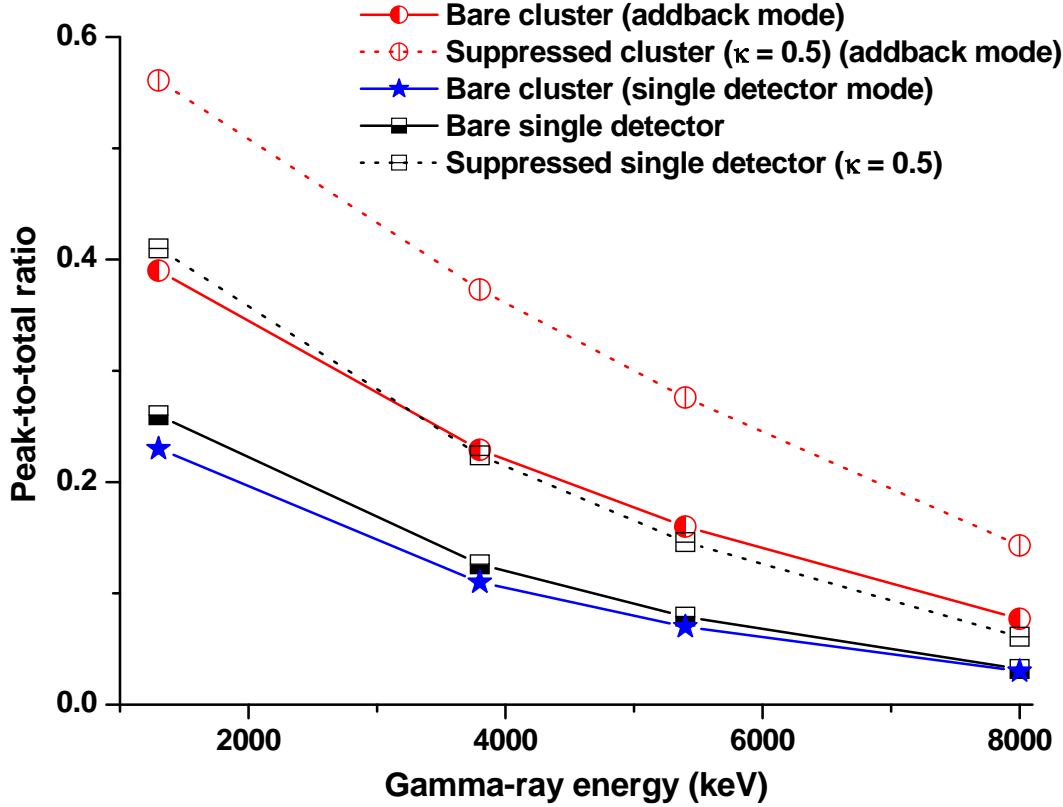}
\caption{Peak-to-total ratio have been plotted as a function of gamma-ray energy for single HPGe and cluster detectors. Results have been shown for both bare and suppressed cases ($\kappa = 0.5$).}\label{fig:new4}
\end{figure}

The variation of peak-to-total ratio as a function of suppression factor $\kappa$, is shown in figure 4. Results have been shown for $\eta$ = 0.1. For low gamma-energy, the variation of peak-to-total is quite smooth. For 8.0 MeV gamma-ray, the variation is smooth up to $\kappa = 0.8$, after which there is a much sharper increase. The dependence of the ratios $R_1$ and $R_2$ as a function of gamma-energy is shown in figure 5. For lower suppression factor $\kappa = 0.1$, $R_1 > 1$ for all energies, which means that decrease in background counts is more due to addback. This situation is different when suppression is better - $R_1 < 1$ for $\kappa = 0.9$. $R_2$ shows little variation with gamma-energy for various suppression cases.

\begin{figure}[htp]
\centering
\includegraphics[totalheight=0.5\textheight,viewport=40 270 780 785,clip]{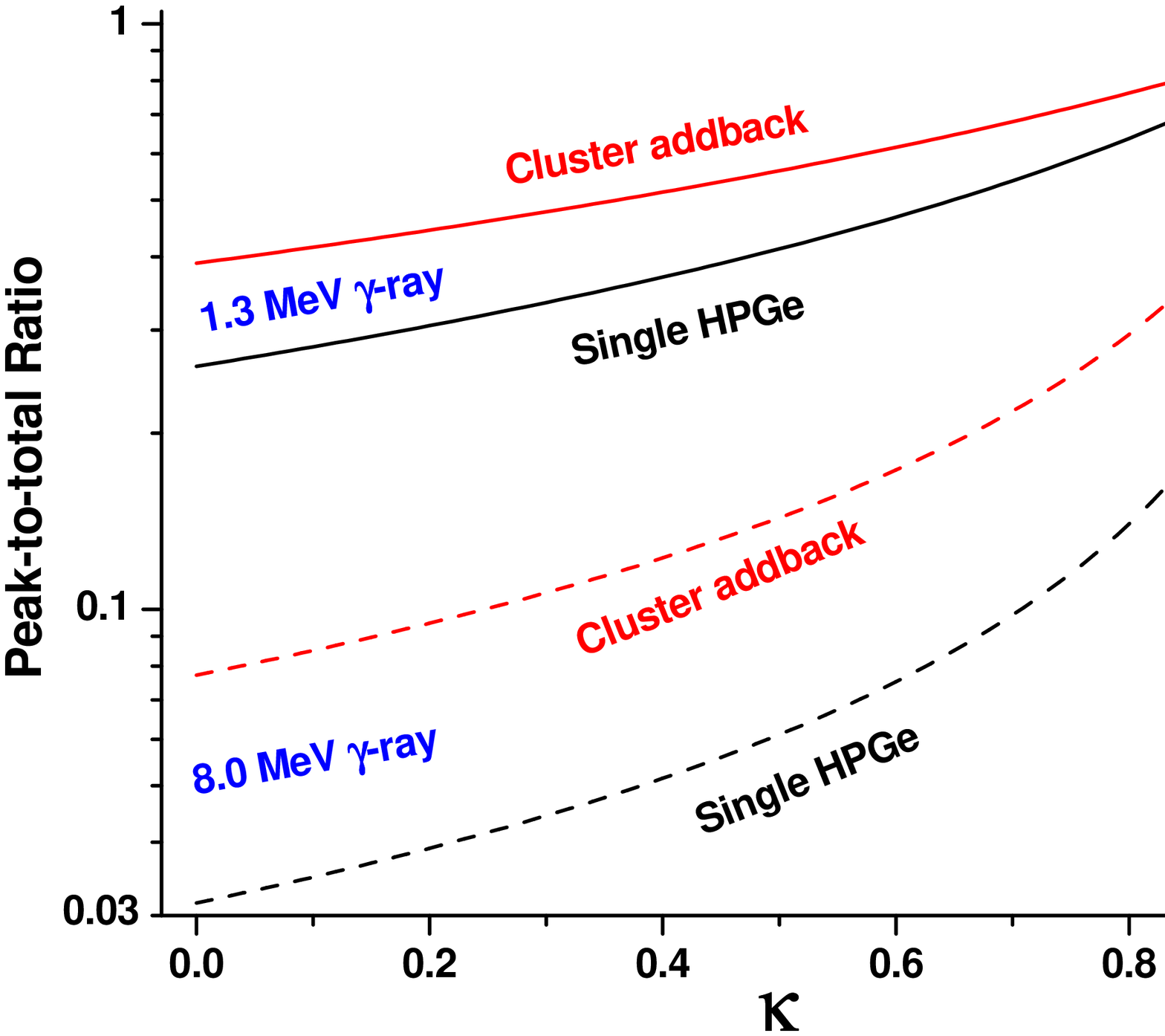}
\caption{Peak-to-total ratio has been plotted as a function of $\kappa$ for the cluster and single HPGe detectors. Results have been shown for two gamma energies.}\label{fig:new6}
\end{figure}

\begin{figure}[htp]
\centering
\includegraphics[totalheight=0.5\textheight,viewport=40 270 780 785,clip]{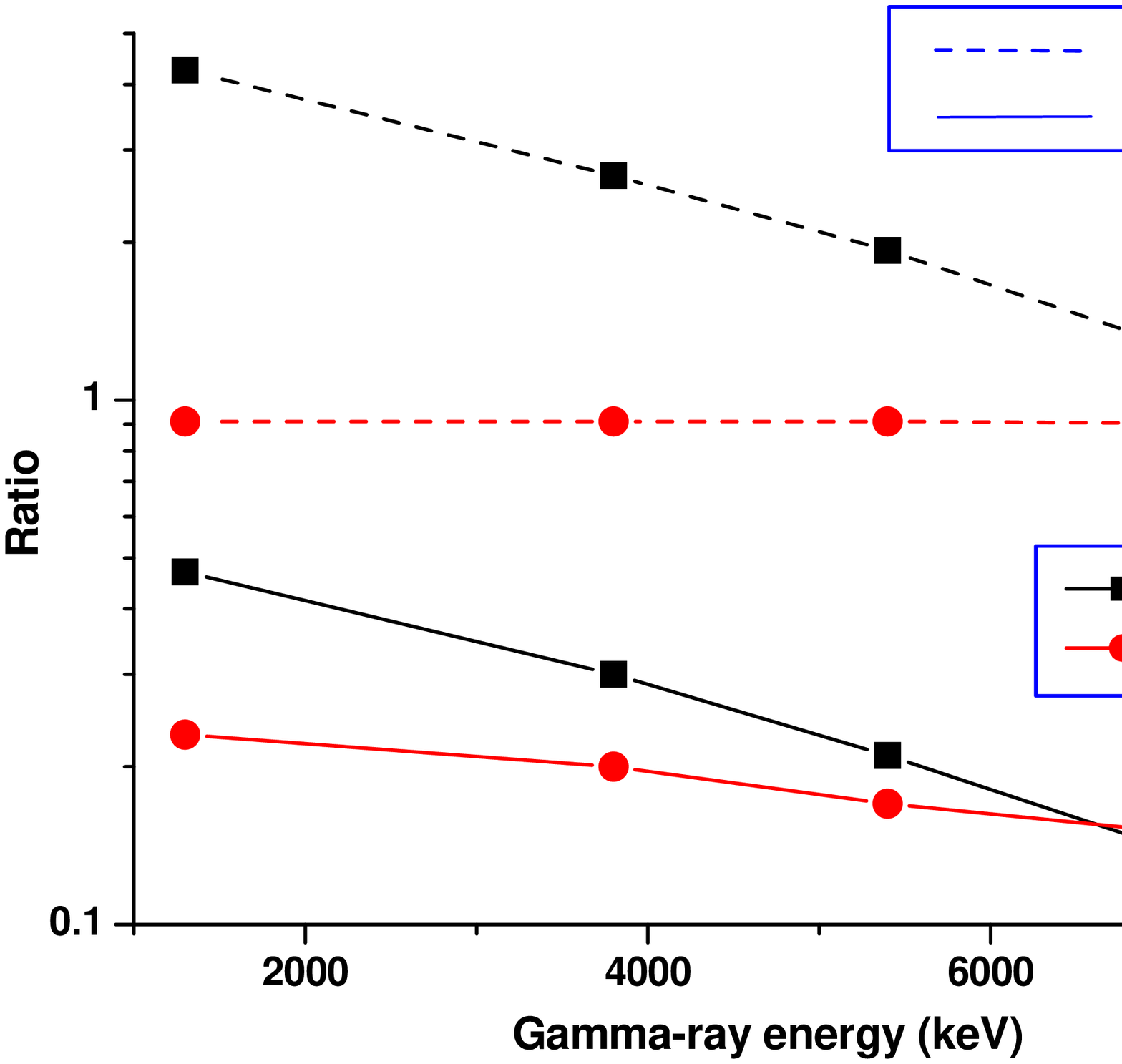}
\caption{Predictions for single and multiple fold events has been plotted along with experimental data for the SPI spectrometer \cite{spi2}.}\label{fig:hitp}
\end{figure}
\newpage

\section{Predictions for various composite detectors using data on cluster detector}

\subsection{Miniball and SPI detectors}

From table 1 of ref. \cite{jinst2}, the respective value of $\rho$ for the cluster, miniball triple configuration and SPI detectors are found to be $\rho_{cluster} = \frac{4}{7}$, $\rho_{mini(3)} = \frac{1}{3}$ and $\rho_{spi} = \frac{14}{19}$, respectively. Thus, we have
\begin{equation}
\alpha_{cluster} = \frac{4}{7}x_1 + \frac{16}{49}x_2  ,  
\end{equation}  
\begin{equation}
\alpha_{mini(3)} = \frac{1}{3}x_1 + \frac{1}{9}x_2  ,  
\end{equation}  
\begin{equation}
\alpha_{spi} = \frac{14}{19}x_1 + \frac{196}{361}x_2    
\end{equation}  
For these detectors, the peak-to-total ratios were calculated at 1332 keV using the experimental data of cluster detector at 1332 keV \cite{jinst2}. In the present work (section 3), we have already predicted the peak-to-total ratios for the cluster detector for energies up to 8 MeV. Following the discussion of section 2.4, we will now calculate the peak-to-total ratio for the miniball three module configuration and SPI detectors. Let us consider these two quantities:
\begin{equation}
\alpha'_{mini(3)} = \frac{1}{3} \times \frac{7}{4} \alpha_{cluster} = \frac{1}{3}x_1 + \frac{1.7}{9}x_2  ,
\end{equation}  
\begin{equation}
\alpha'_{spi} = \frac{14}{19} \times \frac{7}{4} \alpha_{cluster} = \frac{14}{19}x_1 + \frac{152}{361}x_2
\end{equation}  
Comparing with equations 4.4 and 4.5, we have $\alpha'_{mini(3)} - \alpha_{mini(3)} = 0.08x_2$ and $\alpha'_{spi} - \alpha_{spi} = -0.12x_2$. Experimental fold distribution of cluster detector shows that the contribution of triple hit events is $\le 15 \%$ for energies up to 8 MeV (see figure 5 of ref. \cite{wil}). Similarly, for SPI spectrometer this contribution has a maximum value of $\approx 19 \%$ at 8.1 MeV (see table 1 and figure 5 of ref. \cite{spi2}). Thus, for $x_2 = 19 \%$, $\alpha'_{mini(3)} - \alpha_{mini(3)} \approx 2 \%$ and $\alpha'_{spi} - \alpha_{spi} \approx - 2 \%$. Thus, using $\alpha'_{mini(3)} = 0.58 \alpha_{cluster}$, $\alpha'_{spi} = 1.29 \alpha_{cluster}$ and equation 2.7, we could calculate the peak-to-total ratios for these two detectors, and expect a small error for these ratios because of using $\alpha'$ instead of $\alpha$. The results are shown in figure 6. With increasing number of detector modules (say {\it K}), the peak-to-total ratio slowly increases. As an example, for 1.3 MeV $\gamma$-ray - the increase is 65 $\%$ for $K$ = 19 compared to $K$ = 1. At higher energy of 8 MeV, this increase in 200 $\%$. This is because there is a possibility of having more addback contribution with more detector modules. Also, the addback contribution increases with increasing $\gamma$-energy. Thus, we have shown that the present formalism could help in predicting the response of sophisticated detectors.

\begin{figure}[htp]
\centering
\includegraphics[totalheight=0.5\textheight,viewport=40 270 780 785,clip]{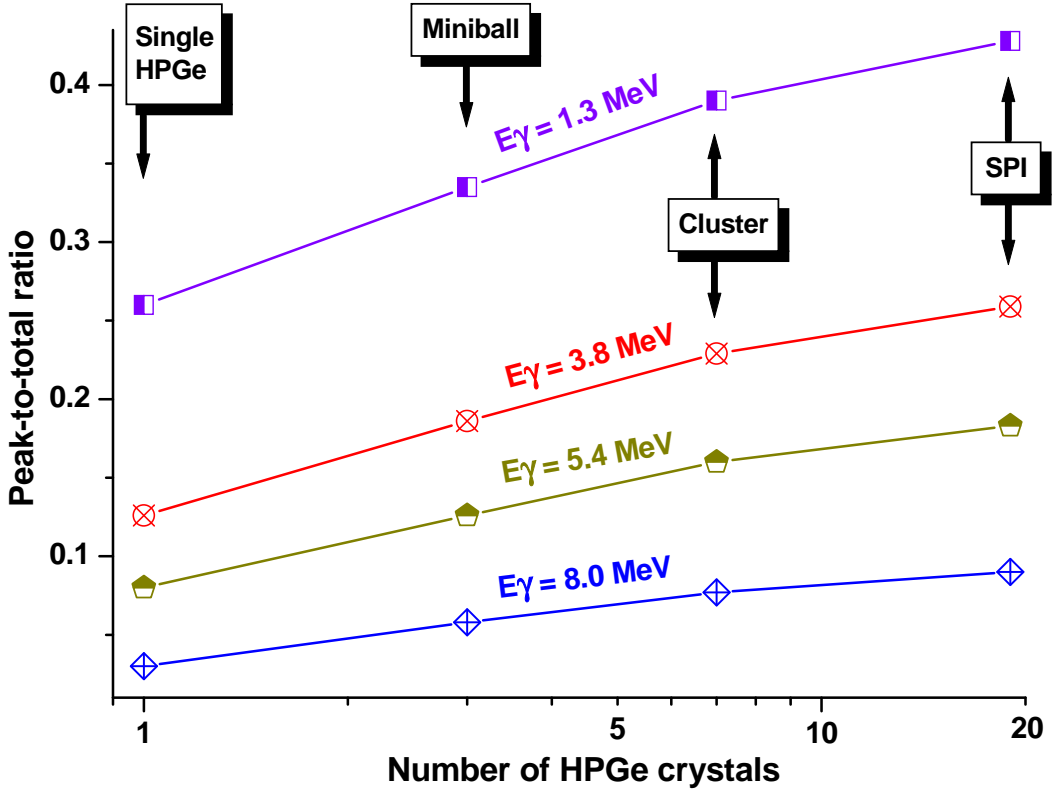}
\caption{Peak-to-total ratio have been plotted as a function of number of detector modules for several bare detectors including single HPGe, cluster miniball three module configuration and SPI detectors. Results have been shown for four energies from 1.3 MeV up to 8 MeV.}\label{fig:new4}
\end{figure}
\newpage

\subsection{Experimental validation for energies up to 8 MeV}

In section 4.3 of ref. \cite{jinst2}, we have shown the comparison of experimental peak-to-total ratio and those obtained from our analysis for energies up to 1.3 MeV. The experimental data of the TIGRESS clover and SPI detectors have been considered. The agreement between the experimental and theoretical peak-to-total ratios shows the experimental validation of our formalism of energies up to 1.3 MeV.

For higher energies say up to 8 MeV, there are no experimental data on peak-to-total ratio of a composite detector. However, data on single and multiple fold distribution of cluster and SPI detectors are present. We will now compare the theoretical and experimental fold distribution for the SPI spectrometer. Experimental value of single and multiple fold probability ($h_s$ and $h_m$, respectively) (data taken from table 1 of \cite{spi2}, $h_s = \frac{\epsilon_{SE}}{\epsilon_{SE} + \epsilon_{ME}}$, $h_m = \frac{\epsilon_{ME}}{\epsilon_{SE} + \epsilon_{ME}}$) and the same calculated theoretically (using equations 2.9 and 2.10) are compared and shown in figure 7. Comparing the plots of $h_s$, we observe a small underestimation of our predictions, which slowly increases with energy - our predicted values are smaller by $1.6 \%$ at 1.3 MeV and $6.0 \%$ at 8.0 MeV. Regarding $h_m$, we observe a small overestimation of our predictions. Nevertheless, in both cases, we observe an agreement (within $6 \%$) between the predictions from our approach and experimental data for SPI spectrometer. As already pointed out in ref. \cite{jinst}, the reason of the small difference between theory and experiment could be due to the different distances at which these detectors are operated. The predictions of the present formalism are based on fact that the source is placed axially at a distance of $\approx$ 25 cm from the detector \cite{ebe,wil}. However, the experimental data of Attie {\it et al.} are based on measurements where the SPI spectrometer is placed at a distance of 125 m from the various calibration sources \cite{spi2}.

\begin{figure}[htp]
\centering
\includegraphics[totalheight=0.5\textheight,viewport=40 270 750 785,clip]{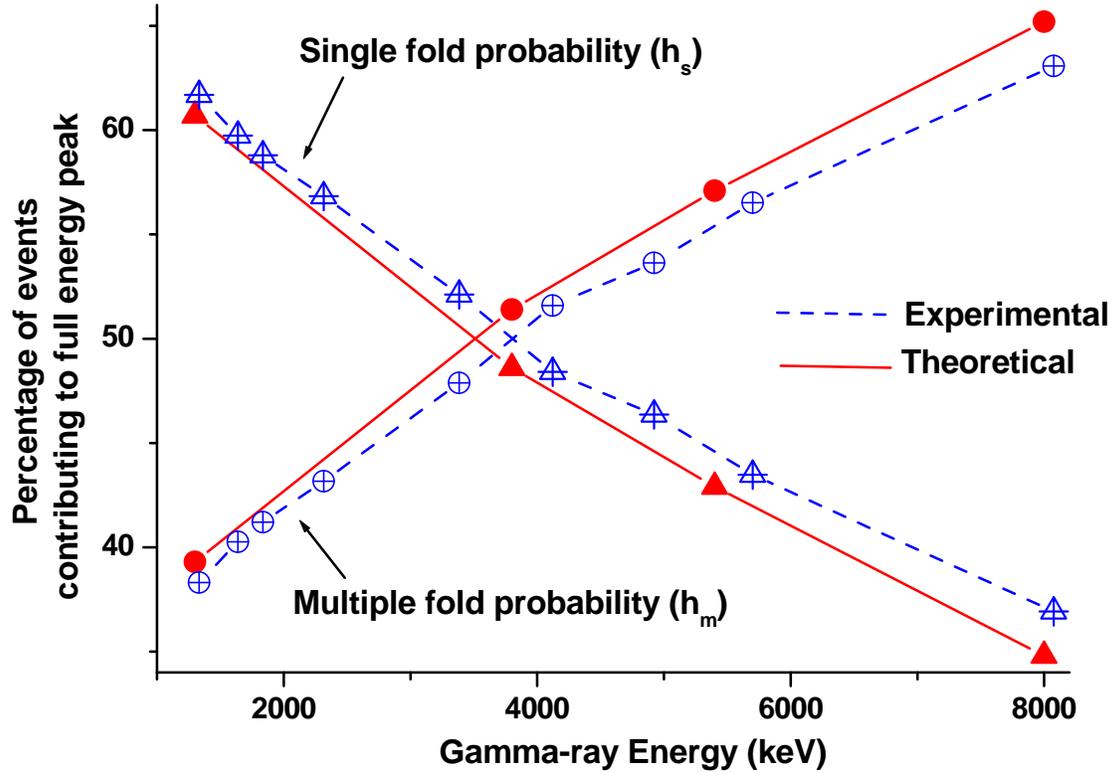}
\caption{Predictions for single and multiple fold events has been plotted along with experimental data for the SPI spectrometer \cite{spi2}.}\label{fig:hitp}
\end{figure}

\subsection{Performance comparison of detectors consisting of several modules of miniball, cluster and SPI detectors}

We will now investigate the fold distribution and peak-to-total ratio of the detectors shown in figure 1 (consisting of several modules of miniball, cluster and SPI detectors) for gamma-energies up to 8 MeV. Our aim is to compare the performance figures for multiple detector units. Note that in ref. \cite{jinst} (see figures 5 and 6(A)) and ref. \cite{jinst2} (see figures 8(A) and 8(B)), the changes in the fold distribution and the peak-to-total ratio of a composite detector as a function of number of detector modules are discussed for 1.3 MeV $\gamma$-ray. 

The first step is the calculation of $\rho$ for each detector. Note that from equations 2.7, 2.8 and 2.15, we observe that a detector system with higher value of $\rho$ will have higher value of peak-to-total ratio and addback factor. Table 2 shows the calculated value of $\rho$ for various detectors of figure 1. The plot of $\rho$ as a function of number of detector units is shown in figure 8(A). Following the procedure shown in section 3 and 4.1, the peak-to-total ratio is calculated for two energies. Results are shown in table 2 and figure 8(B). From table 2 and figures 8(A) and 8(B), we observe:

\begin{table}
\begin{center}
\setlength{\tabcolsep}{0.03in} 
{\caption{\label{tab:table2} Values of $\rho$ ($= \frac{\mu}{3 K}$) for the composite detectors shown in figure 1 are shown along with the calculated value of peak-to-total ratio (PT) for two energies. Here, $2\mu$ is the number of sides which are in contact with each other (see section 3 of ref. \cite{jinst2}).}
\vspace{1.5ex}
\begin{tabular}{|c|p{1.0cm}|p{1.0cm}|p{1.0cm}|c|c|} 
\hline
Detector & {\it K} & $\mu$ & $\rho$ & PT (1.3 MeV) & PT (8.0 MeV) \\
\hline
Ai & 3 & 3 & 0.33 & 0.34 & 0.06 \\
Aii & 9 & 15 & 0.56 & 0.39 & 0.08 \\
Aiii & 21 & 45 & 0.71 & 0.42 & 0.09 \\
\hline
Bi & 7 & 12 & 0.57 & 0.39 & 0.08 \\
Bii & 21 & 45 & 0.71 & 0.42 & 0.09 \\
Biii & 49 & 120 & 0.82 & 0.45 & 0.10 \\
\hline
Ci & 19 & 42 & 0.74 & 0.43 & 0.09 \\
Cii & 57 & 141 & 0.82 & 0.45 & 0.10 \\
Ciii & 133 & 354 & 0.89 & 0.46 & 0.10 \\
\hline
\end{tabular}}
\normalsize
\end{center}
\end{table}

\begin{itemize}

\item $\rho$ increases sharply with increasing value of {\it K}, attaining 0.33 for {\it K} = 3 and 0.57 for {\it K} = 7. Afterwards the increase is much slower. As an example, for {\it K} = 19, $\rho$ = 0.74 and for {\it K} = 133, $\rho$ = 0.89. So, an increase in number of modules by 114 above SPI spectrometer increases $\rho$ by merely 0.15. In other words, compared to the SPI spectrometer, a detector comprising of seven SPI spectrometers has a higher value of $\rho$ by merely 20$\%$. 

\item The increase in peak-to-total ratio (PT) as a function of number of detector modules $\it K$, is much slower compared to that of $\rho$. The rate of increase decreases with higher energies. As an example, at 1.3 MeV, PT changes from 0.34 to 0.46 when {\it K} varies from 3 to 133. At 8.0 MeV, for the same case, PT varies from 0.06 to 0.10. Comparing the cases for $K$ = 19 and 133, we infer that compared to the SPI spectrometer, a detector comprising of seven SPI spectrometers has a higher value of peak-to-total ratio by merely 3$\%$ at 1.3 MeV and 1$\%$ at 8.0 MeV. Since the peak-to-total ratio do not increase significantly after $K$ = 19, so it may not be useful to build larger composite detectors since a small gain in efficiency due to addback after $K$ = 19 comes at the expense of a large number of detector modules. 

\end{itemize}

\begin{figure}[htp]
\centering
\includegraphics[totalheight=0.87\textheight,viewport=20 60 520 800,clip]{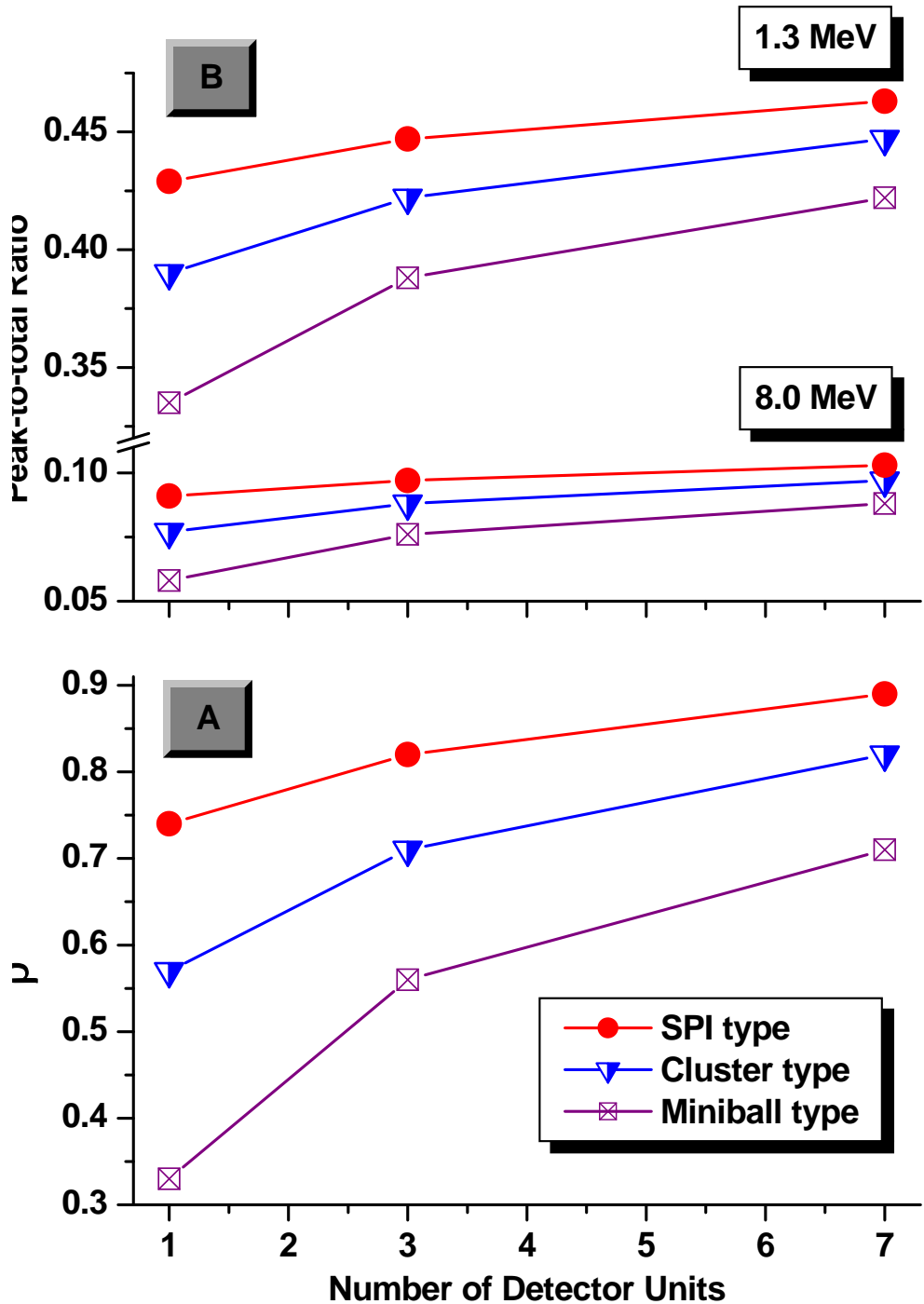}
\caption{Figure A shows the plot of $\rho$ as a function of number of detector (miniball triple cryostat configuration, cluster, SPI) units. The peak-to-total ratio have been plotted in figure B as a function of number of detector units for two energies.}\label{fig:hitp}
\end{figure}

It would be interesting to compare the response of the SPI spectrometer with that of gamma arrays having similar number of detector modules. Since the SPI spectrometer consists of approximately (i) seven times the number of modules of a miniball three detector module cryostat and (ii) three times the number of modules of a cluster detector, let us compare the performance of the SPI spectrometer with two arrays - one comprising of seven miniball detectors (namely array Aiii of figure 1) and another array of three cluster detectors (namely array Bii of figure 1). We know that a large value of $\rho$ means small uncovered lateral surface area and so lesser value of the lateral perimeter. If a hexagonal detector module has lateral perimeter $\it p$, then the lateral perimeter of the SPI spectrometer (Ci of figure 1) and arrays Aiii and Bii of figures 1 are 5.0$\it p$, 6.0$\it p$ and 6.0$\it p$, respectively. Since these two arrays are comprised of same number of modules ($K$ = 21) and they have same lateral perimeter, so from the formalism described in ref. \cite{jinst2}, we can infer that these two arrays will have identical values of peak-to-total ratio. Also, the SPI spectrometer has lower perimeter than these arrays, so we can qualitatively say that the peak-to-total ratio and addback factor of the SPI spectrometer is higher than that of the two arrays. Let us now compare the values of $\rho$. From table 2, we find that $\rho_{Aiii} = \rho_{Bii} = \frac{45}{3\times21} = \frac{5}{7} = 0.71$. Comparing with $\rho_{spi}$ (= 0.74), we infer that the peak-to-total ratio of the SPI spectrometer is slightly better than that of the arrays Aiii and Bii. Thus, having large number of detectors is not crucial for maximizing the peak-to-total ratio but rather their arrangement is important.

The fold distribution of the detectors (of figure 1) as a function of number of detector units is shown in figure 9 for two gamma-energies. For the ease of understanding, the same distribution is shown in figure 10, as a function of number of HPGe detector modules ($K$). At 1.3 MeV, the single fold probability ($h_s$) dominates the distribution for composite detectors up to ${\it K}$ = 133. After decreasing sharply up to $K =$ 19 ($h_s = 61 \%$), $h_s$ decreases at a slow rate (for $K = 133$, $h_s = 56 \%$). Correspondingly, the multiple fold probability ($h_m$), increases sharply up to $K =$ 19 after which the rate of increase becomes much slower. For 8 MeV $\gamma$-ray, the situation is different. In this case, the single fold probability dominates the distribution for composite detector with ${\it K}$ = 3. Correspondingly, the multiple fold probability, increases sharply up to $K =$ 19 after which the rate of increase becomes much slower. As an example, for $K = 19$, $h_m = 65 \%$ and for $K = 133$, $h_m = 69 \%$. Since $h_m$ does not increase significantly after $K$ = 19, so the gain due to the addback process is much lesser if we consider larger composite detectors. Thus, we have shown that the performance of composite detectors (with $K \ge 19$) like the SPI spectrometer does not increase significantly if we consider an array of several SPI spectrometers.

\begin{figure}[htp]
\centering
\includegraphics[totalheight=0.89\textheight,viewport=20 60 520 800,clip]{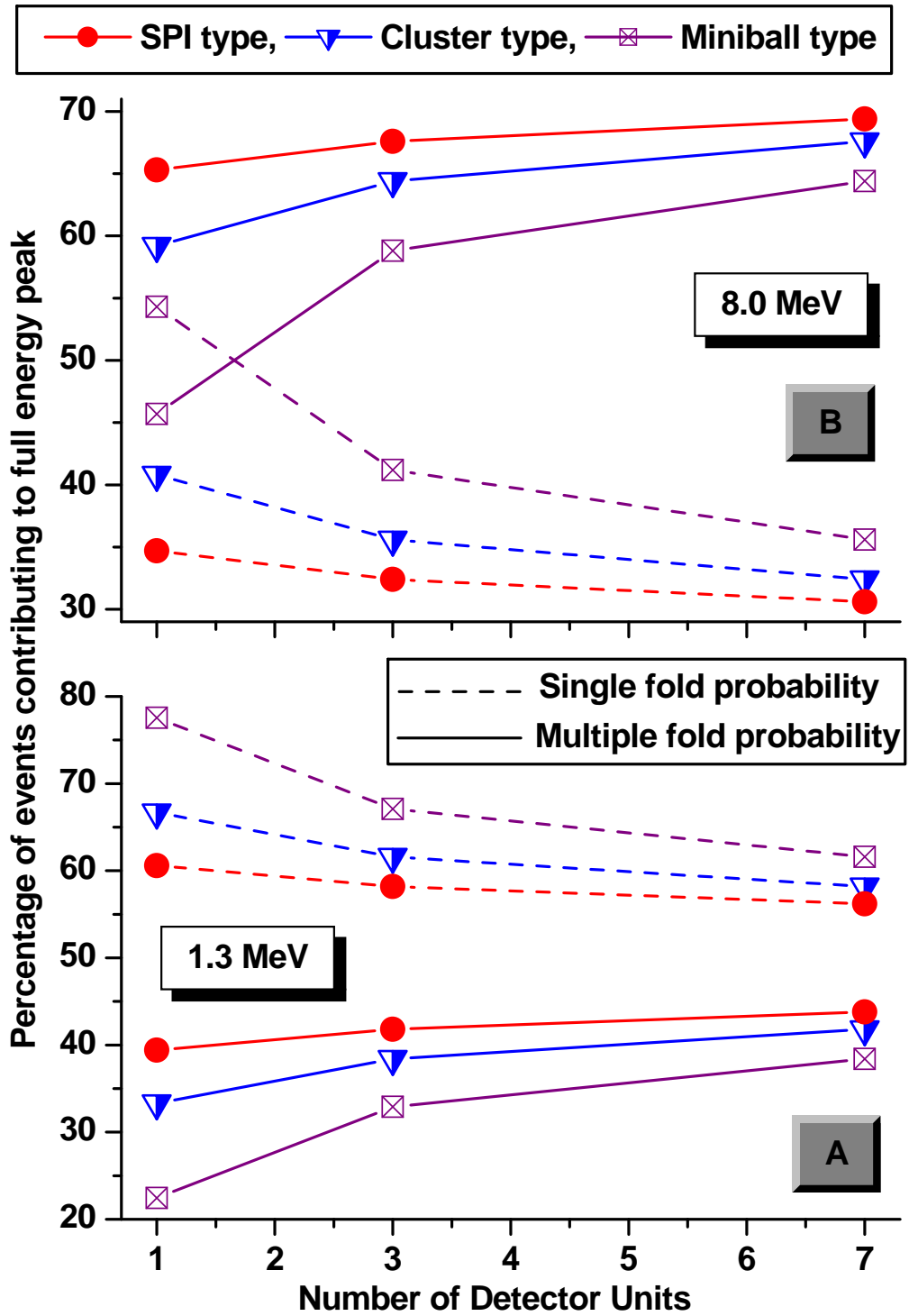}
\caption{The variation of single and multiple fold events are shown for several bare detectors as a function of number of detector units. Figures A and B show the results for 1.3 MeV and 8.0 MeV, respectively.}\label{fig:hitp}
\end{figure}

\begin{figure}[htp]
\centering
\includegraphics[totalheight=0.89\textheight,viewport=20 60 520 800,clip]{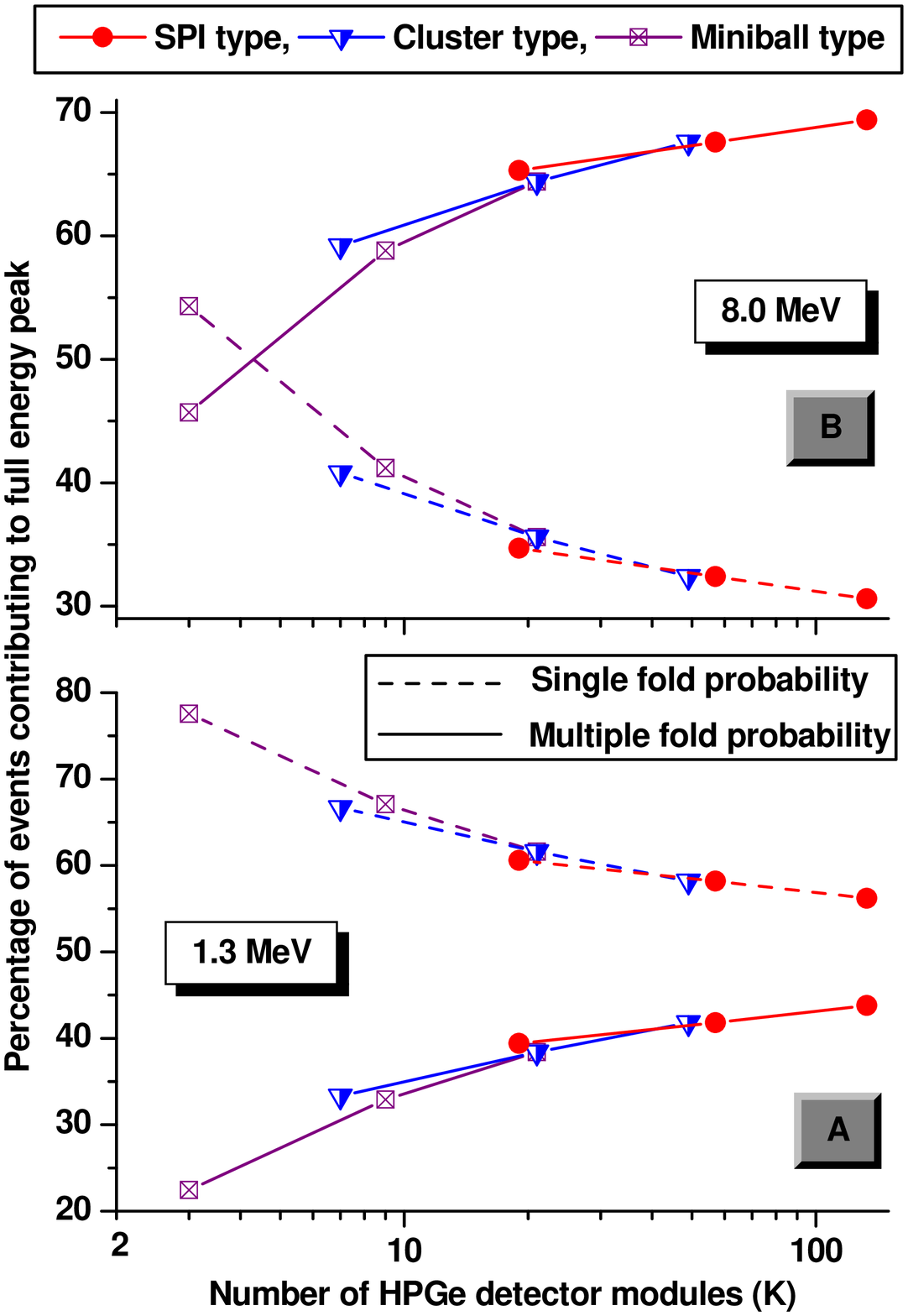}
\caption{The variation of single and multiple fold events are shown for several bare detectors as a function of number of HPGe detector modules. Figures A and B show the results for 1.3 MeV and 8.0 MeV, respectively.}\label{fig:hitp}
\end{figure}

\section{Summary and Conclusion}

Continuing our investigations on the modeling of encapsulated type composite detectors (see R. Kshetri, JINST 2012 7 P04008; ibid., P07006), in the present paper the peak-to-total ratio has been calculated for the first time for energies where direct measurement of peak-to-total ratio is impossible due to absence of a radioactive source having single monoenergetic gamma-ray of that energy. This is an important feature of this formalism. Peak-to-total ratio are calculated for various composite detectors for $\gamma$-energies from 1.3 MeV to 8.0 MeV. It has been observed that both addback mode and active suppression play a crucial role in improving the peak-to-total ratio. We have shown that the predictions of fold distribution match remarkably with experimental data of the SPI spectrometer for energies up to 8 MeV. The case of active suppression has been discussed along with that of bare detector. Performance figures are calculated for both the single detector and addback modes of operation. We have compared the performance of miniball, cluster and SPI detectors with their corresponding arrays for the first time. This work could provide guidance to gamma-ray spectroscopists, in particular, to those that need to understand the responses of their instruments for energies where convenient gamma-ray sources are absent. The present paper is the fifth in the series of papers on composite germanium detectors.

\end{document}